\documentclass[twocolumn,showpacs,eqsecnum,pra]{revtex4}
\usepackage{amsmath}

\usepackage{amssymb}
\usepackage{epsfig}
\begin{document}
\pagenumbering{arabic}

%\draft

\title{Uhlmann fidelity between two-mode Gaussian states}
\author{Paulina Marian$^{1,2}$}
\author{ Tudor A. Marian$^{1}$}
\affiliation{ $^1$Centre for Advanced  Quantum Physics,
University of Bucharest, P.O.Box MG-11,
R-077125 Bucharest-M\u{a}gurele, Romania}
\affiliation{$^2$Department of Physical Chemistry, University of Bucharest,\\
Boulevard Regina Elisabeta 4-12, R-030018  Bucharest, Romania}

%\date{\today}

\begin{abstract}
We analyze the Uhlmann fidelity of a pair of $n$-mode Gaussian states 
of the quantum radiation field. This quantity is shown to be the product 
of an exponential function depending on the relative average displacement 
and a factor fully determined by the symplectic spectrum of the covariance 
matrix of a specific Gaussian state. However, it is difficult to handle 
our general formula unless the Gaussian states commute or at least 
one of them is pure. On the contrary, in the simplest cases $n=1$ and $n=2$, 
it leads to explicit analytic formulae. Our main result is a calculable 
expression of the fidelity of two arbitrary two-mode Gaussian states. 
This can be applied to build reliable measures of quantum correlations
between modes in various branches of quantum physics. 
\end{abstract}
\pacs{03.65.Ta, 03.67.Hk, 42.50.Dv}

%\twocolumn

\maketitle
\section{Introduction.}
One of the most important issues in quantum information is the closeness 
between input and output states of various quantum protocols \cite{BS}.  
When the input state is pure a natural figure of merit of any protocol is the 
input-output quantum-mechanical transition probability. The situation
is more complicated in the mixed-state case. Fortunately, an appropriate 
description in such a case could be connected with a concept introduced 
long ago by Uhlmann \cite{Uhl}. Let us consider two arbitrary states,
$\hat{\rho}^{\prime}$ and $\hat{\rho}^{\prime \prime}$, of a given quantum 
system. When both states are mixed, Uhlmann asserted that a suitable measure
of the likeness between their properties is the maximal quantum-mechanical 
transition probability between their purifications in an enlarged 
Hilbert space \cite{Uhl,Jo}. This is an extended notion of transition 
probability between quantum states which is now called fidelity \cite{Jo} 
and has the intrinsic expression \cite{Uhl}
\begin{equation}
{\cal F}(\hat{\rho}^{\prime}, \hat{\rho}^{\prime \prime})=
\left[ {\rm Tr}\left( \sqrt{\sqrt{\hat{\rho}^{\prime \prime}}\hat{\rho}^{\prime}
\sqrt{\hat{\rho}^{\prime \prime}}} \right) \right]^2.
\label{F} 
\end{equation}
Indeed, when at least one of the quantum states is pure, 
the fidelity\ (\ref{F}) reduces to the familiar overlap between the states:
${\cal F}(\hat{\rho}^{\prime}, \hat{\rho}^{\prime \prime})
={\rm Tr}(\hat{\rho}^{\prime} \hat{\rho}^{\prime \prime})$.
Important properties listed and discussed in Refs.\cite{Uhl,Jo,Bar,Niel,MNMFL,
MPHUZ,BZ} highly recommended fidelity as a measure of distinguishability 
of quantum states. Especially useful is the property put forward 
in Ref. \cite{Bar} that fidelity equals the minimal squared overlap of the probability distributions for the outcomes of any general measurement.
Fidelity-based metrics such as the Bures metric \cite{Bu,Uhl} 
and related ones \cite{Niel,MNMFL,MPHUZ} have proven to be fruitful in quantum optics and quantum information. For instance, the Bures metric provides 
insightful distance-type degrees of entanglement \cite{Pl,PTH}, 
nonclassicality \cite{PTH02}, and polarization \cite{Bj,Luis,OC,IGPT}. 
Moreover, in the framework of the theory of parameter estimation, Braunstein 
and Caves \cite{BC} established a distinguishability metric which
is proportional to the Bures distance between neighboring quantum states. 
Their findings in Ref.\cite{BC} successfully apply in recent developments 
of quantum estimation theory, as shown in the review \cite{Paris}. 
Besides being an unavoidable ingredient in quantum information, fidelity is exploited in various areas of physics where the concept of quantum transition 
probability plays a key role. For instance, in condensed matter physics, 
fidelity is frequently employed as an indicator of criticality in quantum phase transitions \cite{ZPV,Gu}. 

In spite of playing a key role in quantum mechanics due to its operational 
meaning, fidelity between mixed states has a rather limited use so far. This happens because, in general, Uhlmann's expression\ (\ref{F}) is not easy to calculate even on finite-dimensional Hilbert spaces \cite{MNMFL,MPHUZ}. 
Within continuous-variable settings, explicit calculations 
of fidelity were performed for one-mode Gaussian states (GSs) 
in Refs.\cite{Twamley,H}, for some special two-mode GSs in Refs.\cite{PTH}, 
or for commuting two-mode states in Refs.\cite{OC,IGPT}.

In what follows, we concentrate on the GSs of continuous-variable 
systems, which are important for both quantum optics and quantum information processing \cite{FOP}.  From the experimental side, they are quite accessible 
and were effectively obtained with light, Bose-Einstein condensates, 
trapped ions, and Josephson junctions. On the other hand, the GSs 
of the quantum radiation field are important resources in several quantum information protocols \cite{BL,WPGCRSL}. In particular, from the theoretical 
point of view, the two-mode GSs are a perfect test bed for studying entanglement \cite{PTH} or other kinds of correlations between the two modes \cite{PG,AD}. 
Therefore, an exact analytic formula for the fidelity of an arbitrary pair 
of two-mode GSs is highly desirable and its  derivation constitutes a main goal 
of the present paper.  Let us mention that we have previously derived 
the fidelity between two-mode symmetric and squeezed thermal states and used it 
in evaluating a Gaussian degree of entanglement defined with the Bures metric 
\cite{PTH}. Nevertheless, an explicit formula of the fidelity between arbitrary two-mode GSs was not yet found. We acknowledge, however, Ref.\cite{SP} 
as a useful prospect for the multimode Gaussian case.

In order to study the fidelity of two GSs, we find it convenient to introduce 
the operators:
\begin{align}
\hat{\mathcal A}:=\hat{\rho}^{\prime}\hat{\rho}^{\prime \prime}, \quad
\hat{\mathcal B}:=\sqrt{\hat{\rho}^{\prime \prime}}\hat{\rho}^{\prime}
\sqrt{\hat{\rho}^{\prime \prime}}, \quad
\hat{\mathcal C}:=\sqrt{\hat{\rho}^{\prime}}\sqrt{\hat{\rho}^{\prime \prime}},
\label{ABC} 
\end{align} 
with the properties
$\hat{\mathcal B}={\hat{\mathcal C}}^{\dag}\hat{\mathcal C}$ 
and
${\rm Tr}(\hat{\mathcal A})={\rm Tr}(\hat{\mathcal B})\geqq 0$.
When this trace is strictly positive, ${\rm Tr}(\hat{\mathcal A})>0$, 
then a state 
\begin{align}\hat{\rho}_B:=[{\rm Tr}(\hat{\mathcal B})]^{-1}\hat{\mathcal B},
\label{rhoB}
\end{align} 
exists, so that the fidelity\ (\ref{F}) reads:
\begin{align}
{\cal F}(\hat{\rho}^{\prime}, \hat{\rho}^{\prime \prime})=
{\rm Tr}(\hat{\mathcal A})\left[ {\rm Tr}\left( \sqrt{\hat{\rho}_B} \right)
\right]^2.
\label{AB}
\end{align}
Equation\ (\ref{AB}) is the starting point of our explicit evaluation 
of the fidelity between two-mode GSs.
 
From now on, we assume that $\hat{\rho}^{\prime}$ and 
$\hat{\rho}^{\prime \prime}$ are multimode GSs. The plan of the paper is as follows. In Sec. II we recall the phase-space description of the GSs 
and mention Williamson's theorem, which is at the heart of our method. 
Some useful properties of the product of two Hilbert-Schmidt Gaussian operators are collected in Sec. III. Based on the composition rules of such a product, 
we then proceed in Sec.IV to derive the fidelity\ (\ref{AB}) of a pair 
of multimode GSs. We express the second factor in the formula\ (\ref{AB}) 
in terms of the symplectic eigenvalues of the covariance matrix (CM) of the GS 
$\hat{\rho}_B$. We develop our treatment in the multimode framework 
as far as possible. Then we restrict ourselves to the workable case 
of two commuting multimode GSs. In Sec. V, the general expression 
of the fidelity of two GSs enables us to retrieve a previous explicit formula 
for one-mode states, in agreement with Refs.\cite{Twamley,H}. Then we reach 
the main goal of this work: we provide a versatile and workable formula 
for the fidelity between two-mode GSs. An interesting application 
of the two-mode formula concludes Sec. V. The final section deals with a discussion of our analytic results and their potential use in studying 
quantum correlations between modes.
  
\section{ Phase-space description of the Gaussian states.}
An $n$-mode Hilbert-Schmidt operator $\hat{F}$ is defined by its Weyl expansion:
\begin{equation} 
\hat{F}=\frac{1}{(2\pi)^n}\int {\rm d}^{2 n}u \;f(u)\; \hat{D}(-u), 
\label{f}
\end{equation}
where $u$ is a vector in the Euclidian space $\mathbb{R}^{2n}$, the weight 
function $f(u)$ is square-integrable over $\mathbb{R}^{2n}$, and   
$\hat{D}(u)$ is an $n$-mode Weyl displacement operator:
\begin{equation}
\hat{D}(u):=\bigotimes_{j=1}^{n}\hat{D}_j((u)_j).
\label{D}
\end{equation}
Let $\hat{q}_j$ and $\hat{p}_j$ be the canonical quadrature operators
of the $j{\mathrm{th}}$ field mode, and $q_j$ and $p_j$ their eigenvalues.
In Eq.\ (\ref{D}), $(u)_j$ denotes a vector in $\mathbb{R}^{2}$, whose components are precisely $q_j$ and $p_j$:
\begin{align}
(u)_j:=\left(
\begin{matrix}
u_{2j-1}\\ u_{2j}
\end{matrix}
\right)=\left( 
\begin{matrix}
q_j\\ p_j
\end{matrix}
\right), \quad (j=1,2,\dotsc, n).
\label{u_j}
\end{align}
Further,  $\hat{D}_j((u)_j)$ is the corresponding single-mode Weyl displacement operator,
\begin{equation} 
\hat{D}_j((u)_j):=\exp \left[-i\left(q_j\hat{p}_j-p_j\hat{q}_j\right)\right].  
\label{D_j}
\end{equation}

Recall the standard matrix $J$ of the symplectic form on $\mathbb{R}^{2n}$,
which is block-diagonal and skew-symmetric:
\begin{align}
J:=\bigoplus_{k=1}^{n}\,J_k, \quad J_k:=\left( 
\begin{matrix}
0  & 1\\ -1 & 0
\end{matrix}
\right), \;\;\; (k=1,2,\dotsc, n).
\label{J}
\end{align}
Two properties of the Weyl operators\ (\ref{D}) are useful:

$\bullet\;$ composition law:
\begin{equation}
\hat{D}(u)\,\hat{D}(v)=\exp\left(-\frac{i}{2}u^{T}Jv\right)\hat{D}(u+v);
\label{comp}
\end{equation}

$\bullet\;$ orthonormalization relation:
\begin{equation}
{\rm Tr}\left[\hat{D}^{\dag}(u)\hat{D}(v)\right]=(2\pi)^{n}\delta^{(2 n)}(u-v).
\label{ortho}
\end{equation}
For instance, the latter yields the scalar product of two Hilbert-Schmidt 
$n$-mode operators\ (\ref{f}), 
\begin{equation}
{\rm Tr}\left(\hat{F}^{\dag}\hat{G}\right)=\frac{1}{(2\pi)^{n}}
\int{\rm d}^{2 n}u f^{\ast}(u)\,g(u),
\label{HS}
\end{equation}
as well as the trace-formula for a weight function,
\begin{equation}
f(u)={\rm Tr}\left[\hat{D}(u)\hat{F}\right].
\label{wf}
\end{equation}
Note the value $f(0)={\rm Tr}\,(\hat{F})$ and that 
an operator\ (\ref{f}) is self-adjoint iff $f^{\ast}(-u)=f(u)$.

We focus on multimode Hilbert-Schmidt Gaussian operators (HSGOs). 
By definition, the weight function\ (\ref{wf})  of such an operator
is a shifted Gaussian, 
\begin{equation}
f(u)=f(0)\exp\left(-\frac{1}{2}u^{T}{\mathcal F}u-i{\xi}^{T}u\right),
\label{Gwf}
\end{equation}
with a complex shift vector $\xi \in \mathbb{C}^{2n}$ and a symmetric matrix 
${\mathcal F} \in M_{2n}(\mathbb{C})$ whose real part is positive definite:
${\mathcal F}+{\mathcal F}^{\ast}>0$. This implies that the square matrix
${\mathcal F}$ is invertible, so that the Gaussian weight function\ (\ref{Gwf}) 
is integrable \cite{Fol}:
\begin{align}
&\int{\rm d}^{2 n}u\, \exp{\left(-\frac{1}{2}u^{T}{\mathcal F}u-i{\xi}^{T}u\right)}
\notag\\ 
& =(2\pi)^{n}[\det({\mathcal F})]^{-\frac{1}{2}}\exp\left(-\frac{1}{2}
{\xi}^{T}{\mathcal F}^{-1}{\xi}\right), \notag\\
&\left(\Re e\left\{\left[\det\left({\mathcal F}\right)\right]^{\frac{1}{2}}\right\}
>0 \right).
\label{Gint}
\end{align}

A special example of HSGO is the density operator $\hat{\rho}$ of an $n$-mode GS, 
which is positive and of unit trace: ${\rm Tr}(\hat{\rho})=1$. 
Its weight function\ (\ref{wf}), called the characteristic function (CF) 
of the state, is fully determined by the first- and second-order moments 
of all the quadrature operators:
\begin{equation}
\chi (u)=\exp\left[-\frac{1}{2}u^{T}{\mathcal V}\,u
-i\,({\langle{u}\rangle}_{\hat{\rho}})^{T}u \right].
\label{CF}
\end{equation}
In Eq.\ (\ref{CF}), ${\cal V}\in M_{2n}(\mathbb{R})$ denotes the symmetric CM 
of the GS $\hat{\rho}$. It fulfills the concise Robertson-Schr\"odinger 
uncertainty relation, 
\begin{equation}
{\zeta}^{\dag}\left({\cal V}+\frac{i}{2}J \right){\zeta} \geqq 0, \qquad
\left( \zeta \in \mathbb{C}^{2n} \right),
\label{R&S}
\end{equation}
as a necessary and sufficient condition \cite{Simon1,Sc1}. Therefore, 
the CM ${\cal V}$ is positive definite, ${\cal V}>0$, and, according to 
Williamson's theorem \cite{W}, is congruent via a symplectic 
matrix $S \in Sp(2n, \mathbb{R})$ to a diagonal matrix:
\begin{align}
&{\cal V}=S^T\left(\bigoplus_{j=1}^{n}\,{\kappa}_j\,I_j \right)S: \quad
\det({\cal V})=\prod_{j=1}^{n}({\kappa}_j)^2 \; ;  \notag\\
&{\cal V}+\frac{i}{2}J=S^T\left[\bigoplus_{j=1}^{n}\left({\kappa}_j\,I_j 
+\frac{i}{2}J_j\right)\right]S: \notag\\
& \det \left({\cal V}+\frac{i}{2}J \right)
=\prod_{j=1}^{n}\left[({\kappa}_j)^2-\frac{1}{4}\right].
\label{W}
\end{align}
In Eqs.\ (\ref{W}), the positive numbers ${\kappa}_j, \;\; (j=1,2,\dotsc, n),$ 
are the symplectic eigenvalues of the CM ${\cal V}$ \cite{VW}, and all 
the single-mode matrices $I_j$ are equal to the $2\times 2$ identity matrix. 
As a consequence, the $n$-mode Robertson-Schr\"odinger uncertainty relation\ (\ref{R&S}) 
is equivalent to the inequalities 
$\;{\kappa}_j \geqq \frac{1}{2}, \;\; (j=1,2,\dotsc, n).$ 

\section{Product of two Hilbert-Schmidt Gaussian operators.} 
We consider the product $\hat{F}=\hat{F}^{\prime}\hat{F}^{\prime\prime}$
of two $n$-mode HSGOs whose weight functions $f^{\prime}(u)$ and 
$f^{\prime\prime}(u)$ are Gaussians, Eq.\ (\ref{Gwf}). Making use of 
Eqs.\ (\ref{comp}),\ (\ref{ortho}), and\ (\ref{Gint}), one finds that 
the weight function\ (\ref{wf}) of the product $\hat{F}$ has the same form. 
Although we do not list the corresponding composition rules \cite{Fol5} here, 
we just stress that the real part of the resulting symmetric matrix 
${\cal F}\in M_{2n}(\mathbb{C})$ is positive definite in two special cases 
we are interested in:
\begin{enumerate} 
\item Both factors $\hat{F}^{\prime}$ and $\hat{F}^{\prime\prime}$
      are self-adjoint.
\item The factors are adjoints to one another: 
      $\hat{F}^{\prime}=(\hat{F}^{\prime\prime})^{\dag}$.
\end{enumerate}
Hence, in both above-mentioned cases, the product $\hat{F}$ is itself 
an $n$-mode HSGO. In the first case, $\hat{F}$ is not self-adjoint,
unless the two factors commute. In the second one, $\hat{F}$ is a positive 
operator, proportional to the density operator of a GS whose CM is precisely 
${\cal F} \in M_{2n}(\mathbb{R})$. 

We further investigate the product 
$\hat{\mathcal A}:=\hat{\rho}^{\prime}\hat{\rho}^{\prime \prime}$ 
of two\\ $n$-mode GSs. First, making use of Eqs.(2.8)-(2.11), we get its trace:
\begin{align}
&{\rm Tr}\,(\hat{{\mathcal A}})=\left[\det \left({\cal V}^{\prime}
+{\cal V}^{\prime\prime}\right)\right]^{-\frac{1}{2}} \notag\\
&\times \exp{\left[-\frac{1}{2}\left(\delta \langle{u}\rangle \right)^T 
\left({\cal V}^{\prime}+{\cal V}^{\prime\prime}\right)^{-1} 
\delta \langle{u}\rangle \right]}>0. 
\label{TR_HS} 
\end{align}
We have introduced above the relative average $n$-mode displacement 
$\delta \langle{u}\rangle:=\langle{u}\rangle_{{\hat{\rho}}^{\prime}}
-\langle{u}\rangle_{{\hat{\rho}}^{\prime \prime}}.$

 Other properties of the product operator $\hat{\mathcal A}$ are found by evaluating its weight function  \ (\ref{wf}) via Eq.\ (\ref{f}) used for both its factors ${\hat{\rho}}^{\prime}$ and ${\hat{\rho}}^{\prime  \prime }$. After finally performing a routine integral of the type \ (\ref{Gint}) we are left with the Gaussian weight function $f_A(u)$,
Eq.\ (\ref{Gwf}), which is determined by the following composition rules:
\begin{align}
& {\cal F}_A=-\frac{i}{2}J+\left({\cal V}^{\prime \prime}+\frac{i}{2}J \right)
\left({\cal V}^{\prime}+{\cal V}^{\prime \prime}\right)^{-1}
\left({\cal V}^{\prime}+\frac{i}{2}J \right)\, ;\notag \\
& {\xi}_A={\langle u \rangle}_{\hat{\rho}^{\prime}}-\left({\cal V}^{\prime}
-\frac{i}{2}J \right)\left({\cal V}^{\prime}+{\cal V}^{\prime\prime}\right)^{-1}
\delta \langle{u}\rangle. 
\label{rhorho}
\end{align}
Similarly, we employ the Hilbert-Schmidt scalar product\ (\ref{HS}) 
in conjunction with the Gaussian integral\ (\ref{Gint})
in order to get the formula
\begin{equation}
{\rm Tr}\left({\hat{\mathcal A}}^2\right)
=\left[{\rm Tr}\,(\hat{\mathcal A})\right]^{2}
2^{-n}\left[\det \left({\cal F}_A \right)\right]^{-\frac{1}{2}}. 
\label{Tr(A^2)}
\end{equation}
An essential non-trivial step is that the determinant of the matrix 
${\cal F}_A$, Eq.\ (\ref{rhorho}), always factors as follows:
\begin{equation}
\det\,({\mathcal F}_A)=\frac{\det \left[(J{\mathcal V}^{\prime})
\,(J{\mathcal V}^{\prime \prime})-\frac{1}{4}I\right]}
{\det \left({\mathcal V}^{\prime}+{\mathcal V}^{\prime \prime}\right)}\,.
\label{detF_A}
\end{equation}
 The determinant \ (\ref{detF_A}) easily emerges 
when arranging the matrix ${\mathcal F}_{A}$ in a product form.
On the other hand, Eq.\ (\ref{rhorho}) implies the identity
\begin{equation}
\det \left({\cal F}_A+\frac{i}{2}\,J\right)=
\frac{\det \left({\cal V}^{\prime }+\frac{i}{2}\,J \right)
\det \left({\cal V}^{\prime \prime}+\frac{i}{2}\,J\right)}
{\det \left({\cal V}^{\prime}+{\cal V}^{\prime \prime}\right)}.
\label{detF_A+J}
\end{equation}
Both the above determinants\ (\ref{detF_A}) and\ (\ref{detF_A+J}) 
are manifestly symmetric with respect to the GSs $\hat{\rho}^{\prime}$ 
and $\hat{\rho}^{\prime \prime}$.

We exploit the composition rules\ (\ref{TR_HS}) and \ (\ref{rhorho}) in three special situations.
First, when the GSs coincide, 
$\hat{\rho}^{\prime}=\hat{\rho}^{\prime \prime}=:\hat{\rho}$, then 
${\cal F}_A={\cal V}_B=:{\cal V}_A$. Hence, Eqs.\ (\ref{TR_HS}) 
and\ (\ref{rhorho}) read: 
\begin{align}
 {\rm Tr}\left({\hat{\rho}}^2\right)=2^{-n}\left[\det({\cal V})
\right]^{-\frac{1}{2}} \leqq 1\, ; \label{a1}\end{align}
\begin{align} {\cal V}_A=\frac{1}{2}\left({\cal V}-\frac{1}{4}J{\cal V}^{-1}J \right),\;\;\,  
\qquad {\xi}_A=\langle{u}\rangle_{\hat{\rho}}.
\label{rho^2}
\end{align}
Second, Eqs.\ (\ref{a1}), \ (\ref{rho^2}) and\ (\ref{W}) enable us to write down two conditions,
both of them necessary and sufficient in order that an $n$-mode GS 
$\hat{\sigma}$ is pure. They are expressed in terms of its CM ${\cal V}$, 
as follows: 
\begin{itemize} 
\item Scalar condition: 
\begin{align}
& {\rm Tr}\left({\hat{\sigma}}^2\right)=1 \iff \det({\cal V})=2^{-2n} \iff \notag\\
& \iff {\kappa}_j =\frac{1}{2}\, , \;\; (j=1,2,\dotsc, n);
\label{pure1}
\end{align}
\item Matrix condition:
\begin{equation}
{\cal V}_A={\cal V} \iff \left(J{\cal V}\right)^2=-\frac{1}{4}\,I.
\label{pure2}
\end{equation}
\end{itemize}
Third, the square root $\sqrt{\hat{\rho}}$ of a given $n$-mode GS 
$\hat{\rho}$ with the CF\ (\ref{CF}) is proportional to an equally displaced 
$n$-mode GS: its CM $\tilde{\cal V}$ determines the proportionality factor 
\begin{equation}{\rm Tr}\left(\sqrt{\hat{\rho}}\right)=\sqrt[4]{\det(2\tilde{\cal V})},
\label{12Tr} 
\end{equation}
and satisfies Eq.\ (\ref{rho^2}):
\begin{equation}
{\cal V}=\frac{1}{2}\left(\tilde{\cal V}
-\frac{1}{4}J{\tilde{\cal V}}^{-1}J \right).
\label{tildeV}
\end{equation}
Accordingly, its symplectic eigenvalues ${\tilde{\kappa}}_j$ are determined 
by those of the CM ${\cal V}$, hereafter denoted ${{\kappa}}_j$:
\begin{equation}
{\tilde{\kappa}}_j={\kappa}_j+\sqrt{{\left({\kappa}_j \right)}^2-\frac{1}{4}}, 
\qquad (j=1,2,\dotsc, n).
\label{tildekappa}
\end{equation}

\section{Fidelity of two multimode Gaussian states.}
We successively apply Eqs.\ (\ref{12Tr}), \ (\ref{W}), and\ (\ref{tildekappa}) 
to the GS $\hat{\rho}_B$, Eq.\ (\ref{rhoB}), for getting a general expression 
of the fidelity\ (\ref{AB}):
\begin{align}
{\cal F}(\hat{\rho}^{\prime}, \hat{\rho}^{\prime \prime})
& ={\rm Tr}(\hat{\rho}^{\prime}\,\hat{\rho}^{\prime \prime}) \notag \\
& \times 2^{n}\prod_{j=1}^{n}\left\{({\kappa}_B)_j
+\sqrt{\left[({\kappa}_B)_j\right]^2-\frac{1}{4}}\right\}.
\label{nF} 
\end{align}
In Eq.\ (\ref{nF}), the set $\left\{\,({\kappa}_B)_j, \;\; 
(j=1,2,\dotsc, n)\right\}$ is the symplectic spectrum of the CM ${\cal V}_B$ 
of the $n$-mode GS $\hat{\rho}_B$. 
Some expected properties of the fidelity can be read on the above expression.
For instance, let us note the inequalities
\begin{equation}
{\cal F}({\hat{\rho}}^{\prime}, {\hat{\rho}}^{\prime \prime})\geqq
{\rm Tr}({\hat{\rho}}^{\prime} {\hat{\rho}}^{\prime \prime})>0,
\label{ineq}
\end{equation}
with saturation of the first one when at least one of the GSs 
$\hat{\rho}^{\prime}$ and $\hat{\rho}^{\prime \prime}$ is pure.
The second inequality is displayed by Eq.\ (\ref{TR_HS}): 
the Hilbert-Schmidt scalar product of any pair of $n$-mode GSs is strictly positive. Another property of the fidelity\ (\ref{nF}) is its symplectic invariance:
\begin{equation}
{\cal F}\left(\hat{U}(S){\hat{\rho}}^{\prime}\hat{U}^{\dag}(S),\, 
\hat{U}(S){\hat{\rho}}^{\prime \prime}\hat{U}^{\dag}(S) \right)
={\cal F}(\hat{\rho}^{\prime}, \hat{\rho}^{\prime \prime}), 
\label{invF}
\end{equation}
where $S \in Sp(2n, \mathbb{R})$ and $\hat{U}(S)$ are the unitary operators 
of the metaplectic representation on the Hilbert space of the $n$-mode states.

There is a special case when the multimode fidelity\ (\ref{nF}) can be readily 
evaluated, namely, that of commuting GSs. As a prototype of such states, 
we mention the $n$-mode thermal states. We concentrate on a pair of commuting 
$n$-mode GSs, $\hat{\rho}^{\prime}$ and $\hat{\rho}^{\prime \prime}$, whose CMs ${\cal V}^{\prime}$ and ${\cal V}^{\prime \prime}$ have the symplectic 
spectra $\{{\kappa}_j^{\prime}\}$ and $\{{\kappa}_j^{\prime \prime}\}$, 
respectively. The commutation relation 
$[\hat{\rho}^{\prime},\, \hat{\rho}^{\prime \prime}]=\hat{0}$ implies, 
via the resulting equality ${\cal V}_B={\cal F}_A$, that the diagonalizable 
matrices $J{\cal V}^{\prime}$ and $J{\cal V}^{\prime \prime}$ commute, 
so that they have a complete system of common eigenvectors. Since the matrix 
$J{\cal V}_B$ has the same eigenvectors, we find the composition law of the symplectic eigenvalues:
\begin{equation}
({\kappa}_B)_j=\frac{{\kappa}_j^{\prime}\kappa_j^{\prime \prime}+\frac{1}{4}}
{{\kappa}_j^{\prime}+\kappa_j^{\prime \prime}}, \qquad
(j=1,2,\dotsc, n).
\label{kappa} 
\end{equation} 
Accordingly, the fidelity formula\ (\ref{nF}) simplifies to: 
\begin{align}
& {\cal F}(\hat{\rho}^{\prime}, \hat{\rho}^{\prime \prime})
= \prod_{j=1}^{n} \frac{2}{\left( {\kappa}_j^{\prime}
+{\kappa}_j^{\prime \prime} \right)^2}  \notag \\
& \times\left\{ {\kappa}_j^{\prime}\kappa_j^{\prime \prime}+\frac{1}{4} 
+\sqrt{\left[ \left( {\kappa}_j^{\prime} \right)^2
-\frac{1}{4} \right]
\left[ \left( {\kappa}_j^{\prime \prime} \right)^2-\frac{1}{4} \right]}
\right\}. 
\label{nFcom} 
\end{align}
Equation\ (\ref{nFcom}) displays two additional properties of the fidelity,
namely, its symmetry,
${\cal F}(\hat{\rho}^{\prime  \prime}, \hat{\rho}^{\prime})
={\cal F}(\hat{\rho}^{\prime}, \hat{\rho}^{\prime \prime}),$
and the saturable inequality
\begin{equation}
{\cal F}(\hat{\rho}^{\prime}, \hat{\rho}^{\prime \prime})\leqq 1:
\qquad {\cal F}(\hat{\rho}^{\prime}, \hat{\rho}^{\prime \prime})=1
\;\;\; \text{iff} \;\;\; \hat{\rho}^{\prime \prime}=\hat{\rho}^{\prime}.
\label{F<1} 
\end{equation}

We come back to the general case of two non-commuting $n$-mode GSs: writing 
an explicit formula of their fidelity\ (\ref{nF}) seems to be a hard task. 
Indeed, evaluation of the  CM ${\cal V}_B$ of the $n$-mode GS 
$\hat{\rho}_B$, Eq.\ (\ref{rhoB}), requires a three-time application of the composition rule \ (\ref{rhorho}) and therefore the CM
${\cal V}_B \in M_{2n}(\mathbb{R})$ has a complicated structure. 
However, we succeeded to calculate its symplectic invariants 
$\det\,({\cal V}_B)$ and $\det\left({\cal V}_B+\frac{i}{2}J \right)$. Indeed, taking advantage once again of Eqs.\ (\ref{HS}) and\ (\ref{Gint}), we find a formula which is similar to Eq.\ (\ref{Tr(A^2)}):
\begin{equation}
{\rm Tr}\left({\hat{\mathcal B}}^2\right)
=\left[{\rm Tr}\,(\hat{\mathcal B})\right]^{2}
2^{-n}\left[\det \left({\cal V}_B \right)\right]^{-\frac{1}{2}}. 
\label{Tr(B^2)}
\end{equation}
Their comparison gives, via Eq.\ (\ref{ABC}), the expression\ (\ref{detF_A}): 
\begin{equation}
\det\,({\cal V}_B)=\frac{\det \left[(J{\cal V}^{\prime})
\,(J{\cal V}^{\prime \prime})-\frac{1}{4}I\right]}
{\det \left({\cal V}^{\prime}+{\cal V}^{\prime \prime}\right)}\geqq 2^{-2n}.
\label{detV_B}
\end{equation}
On the other hand, a rather involved calculation makes use of 
Eqs.\ (\ref{rhorho}) and\ (\ref{tildeV}) to yield the determinant
\begin{equation}
\det \left({\cal V}_B+\frac{i}{2}\,J\right)=
\frac{\det \left({\cal V}^{\prime }+\frac{i}{2}\,J \right)
\det \left({\cal V}^{\prime \prime}+\frac{i}{2}\,J\right)}
{\det \left({\cal V}^{\prime}+{\cal V}^{\prime \prime}\right)},
\label{detV_B+J}
\end{equation}
which coincides with the non-negative determinant\ (\ref{detF_A+J}). 

When at least one of the GSs $\hat{\rho}^{\prime}$ and 
$\hat{\rho}^{\prime \prime}$, say $\hat{\rho}^{\prime}$, is pure, 
then insertion of the appropriate conditions\ (\ref{pure1}) and\ (\ref{pure2}) 
into Eq.\ (\ref{detV_B}) leads to the equality $\det\,({\cal V}_B)=2^{-2n}$,
showing that the GS $\hat{\rho}_B$ is also pure.

For later convenience, let us introduce the following notations:
\begin{align}
& \Delta:=\det \left({\cal V}^{\prime}+{\cal V}^{\prime \prime}\right) \geqq 1;
\notag \\
& \Gamma:=2^{2n}\det \left[(J{\cal V}^{\prime})
\,(J{\cal V}^{\prime \prime})-\frac{1}{4}I\right] \geqq \Delta \, ; \notag \\
& \Lambda:=2^{2n}\det \left({\cal V}^{\prime }+\frac{i}{2}\,J \right)
\det \left({\cal V}^{\prime \prime}+\frac{i}{2}\,J\right) \geqq 0.
\label{Delta} 
\end{align}
The inequality $\Delta \geqq 1$ originates in the feature of the positive 
definite matrix $\check{\cal V}:=\frac{1}{2}\left({\cal V}^{\prime}
+{\cal V}^{\prime \prime}\right)$ of being the CM of an $n$-mode GS determined
up to its average displacement.
Let us remark that our main results in this section are the explicit 
expressions of the symplectic invariants\ (\ref{detV_B}) and 
\ (\ref{detV_B+J}) for GSs with any $n$. However, only for one- and two-mode states these invariants are sufficient to get analytic formulae 
for the fidelity via Eq.\ (\ref{nF}).  

\section{Explicit results}
\subsection{One-mode Gaussian states.} 
For the sake of completitude we begin with the one-mode case. The symplectic spectrum of the CM ${\cal V}_B$ consists of a single eigenvalue,  
$${\kappa}_B=\frac{1}{2}\sqrt{\frac{\Lambda}{\Delta}+1}.$$ 
According to Eqs.\ (\ref{rhorho}) and\ (\ref{Delta}), the fidelity\ (\ref{nF}) 
reads:
\begin{align}
{\cal F}({\hat{\rho}}^{\prime}, {\hat{\rho}}^{\prime \prime})
& =\exp{\left[-\frac{1}{2}\left(\delta \langle{u}\rangle \right)^T 
\left({\cal V}^{\prime}+{\cal V}^{\prime\prime}\right)^{-1} 
\delta \langle{u}\rangle \right]} \notag \\
& \times\left( \sqrt{\Delta+\Lambda}-\sqrt{\Lambda} \right)^{-1}.
\label{1F}
\end{align}
The symmetric formula\ (\ref{1F}) was derived and analyzed gradually long ago 
in Refs. \cite{H}. In addition, by using the structure of the one-mode GSs, 
which are displaced squeezed thermal states \cite{PT}, we succeeded to prove 
explicitly in the present work the property\ (\ref{F<1}) 
of their fidelity\ (\ref{1F}).

\subsection{Two-mode Gaussian states.}
The CM ${\cal V}_B$ has two symplectic eigenvalues, $({\kappa}_B)_1$ and
$({\kappa}_B)_2$. Their squares are the roots of a monic quadratic binomial
whose coefficients are expressed in terms of the determinants\ (\ref{Delta}).
A straightforward calculation exploiting Eqs.\ (\ref{rhorho}) and\ (\ref{nF}) 
leads to the symmetric formula
\begin{align}
& {\cal F}({\hat{\rho}}^{\prime}, {\hat{\rho}}^{\prime \prime}) 
=\exp{\left[-\frac{1}{2}\left(\delta \langle{u}\rangle \right)^T 
\left({\cal V}^{\prime}+{\cal V}^{\prime\prime}\right)^{-1} 
\delta \langle{u}\rangle \right]} \notag \\
& \times\left[ \left( \sqrt{\Gamma}+\sqrt{\Lambda} \right)
-\sqrt{\left( \sqrt{\Gamma}+\sqrt{\Lambda} \right)^2-\Delta} \right]^{-1}.
\label{2F}
\end{align}
Previous results reported in Refs.\cite{PTH} for special states (squeezed thermal and symmetric) are seen to be particular cases of Eq.\ (\ref{2F}).

The determinants\ (\ref{Delta}) of  $4\times4$  matrices occurring 
in Eq.\ (\ref{2F}) are easily computable. As a salient example, 
we evaluate them for a pair of undisplaced two-mode GSs whose CMs 
${\cal V}^{\prime}$ and ${\cal V}^{\prime \prime}$ are both in an unscaled standard form. This means that they are partitioned into $2\times 2$ diagonal submatrices of the type
\begin{align}
{\cal V}=\left(
\begin{matrix} 
b_{1}I       & {\mathcal C}\\ 
{\mathcal C} & b_{2}I 
\end{matrix}
\right):\quad
I=\left(
\begin{matrix}
1 & 0\\
0 & 1
\end{matrix}
\right),\quad
{\mathcal C}=\left(
\begin{matrix} 
c & 0\\ 
0 & d 
\end{matrix}
\right). \; 
\label{standard}
\end{align}
Here $b_{1}\geqq \frac{1}{2},\; b_{2}\geqq \frac{1}{2},\; c\geqq |d|.$
A matrix\ (\ref{standard}) has two symplectic invariants:
\begin{align}
& \det({\cal V})=\left( b_{1} b_{2}-c^2 \right)\left( b_{1} b_{2}-d^2 \right)>0, 
\notag \\
& \det \left({\cal V}+\frac{i}{2}J \right)=\det({\cal V})
-\frac{1}{4}\left( b_{1}^2+b_{2}^2+2cd \right)+\frac{1}{16}\geqq 0.
\label{Sp4}
\end{align}
We denote $b^{\prime}_{1}, b^{\prime}_{2}, c^{\prime}, d^{\prime}$ and 
$b^{\prime \prime}_{1}, b^{\prime \prime}_{2}, c^{\prime \prime},
d^{\prime \prime}$ the standard-form parameters of the CMs ${\cal V}^{\prime}$
and ${\cal V}^{\prime \prime}$, respectively. Then, Eqs.\ (\ref{Delta}), 
\ (\ref{standard}), and\ (\ref{Sp4}) allow us to factor both determinants 
$\Delta$ and $\Gamma$ and to visualize the symplectic invariant $\Lambda$:  
\begin{align} 
\Delta=\left[ \left( b^{\prime}_{1}+b^{\prime \prime}_{1} \right)
\left( b^{\prime}_{2}+b^{\prime \prime}_{2} \right)
-\left( c^{\prime}+c^{\prime \prime} \right)^2 \right] \notag \\ 
\times \left[ \left( b^{\prime}_{1}+b^{\prime \prime}_{1} \right)
\left( b^{\prime}_{2}+b^{\prime \prime}_{2} \right)
-\left( d^{\prime}+d^{\prime \prime} \right)^2 \right]; \notag \\ 
\Gamma =16\left\{ \left[ b^{\prime}_{1}b^{\prime}_{2}-\left( d^{\prime} 
\right)^2 \right]\left[ b^{\prime \prime}_{1}b^{\prime \prime}_{2}
-\left( c^{\prime \prime} \right)^2 \right] \right. \notag \\
\left. +\frac{1}{4}\left( b^{\prime}_{1}b^{\prime \prime}_{1}
+b^{\prime}_{2}b^{\prime \prime}_{2}+2d^{\prime}c^{\prime \prime} \right)
+\frac{1}{16} \right\} \notag \\
\times \left\{ \left[ b^{\prime}_{1}b^{\prime}_{2}-\left( c^{\prime} 
\right)^2 \right]\left[ b^{\prime \prime}_{1}b^{\prime \prime}_{2}
-\left( d^{\prime \prime} \right)^2 \right] \right. \notag \\
\left. +\frac{1}{4}\left( b^{\prime}_{1}b^{\prime \prime}_{1}
+b^{\prime}_{2}b^{\prime \prime}_{2}+2c^{\prime}d^{\prime \prime} \right)
+\frac{1}{16} \right\}; \notag \\ 
\Lambda=16\left\{\det({\cal V}^{\prime})
-\frac{1}{4}\left[ (b_{1}^{\prime})^2+(b_{2}^{\prime})^2
+2c^{\prime}d^{\prime} \right]+\frac{1}{16} \right\} \notag \\
\times \left\{\det({\cal V}^{\prime \prime})
-\frac{1}{4}\left[ (b_{1}^{\prime \prime})^2+(b_{2}^{\prime \prime})^2
+2c^{\prime \prime}d^{\prime \prime} \right]+\frac{1}{16} \right\}.
\label{Delta2}
\end{align}
We are left to substitute the formulae\ (\ref{Delta2}) into Eq.\ (\ref{2F}) written for undisplaced two-mode GSs. 

\section{Conclusions}
To sum up, in this work we have tackled a long-standing problem in quantum
information with continuous variables: the Uhlmann fidelity of two 
multimode GSs. Our simple phase-space approach led to the important 
formulae\ (\ref{Tr(A^2)})--\ (\ref{detF_A+J}) and
\ (\ref{Tr(B^2)})--\ (\ref{detV_B+J}) which are valid for any number of modes 
and could be useful for future research. However, we stress that an analytic 
result for $n$-mode GSs seems hard to be found when $n \geqq 3:$ 
indeed, one then needs to solve a higher-degree characteristic equation
that cannot be easily obtained. 
 
While in the single-mode case an explicit formula was found many years ago, 
we derived here a computable analytic expression in the two-mode case, 
Eq.\ (\ref{2F}). We expect that this formula  will open a productive research 
in quantifying bipartite quantum correlations (entanglement and discord) 
in the Gaussian scenario.  An example at hand is the already performed 
evaluation of a degree of entanglement for some important classes of GSs 
(squeezed thermal states, symmetric states) in Refs.\cite{PTH}. The analysis 
of an arbitrary two-mode GS requires our present result. Use of fidelity 
in evaluating the Gaussian geometric discord defined with the Bures metric is currently one of our main interests.  Note also that fidelity in mixed-state estimation or reconstruction  could be a potential important tool. 
It appears that applications of the fidelity to the emerging field 
of quantum information and to other interesting areas of quantum physics 
are far from being exhausted.

{\bf Acknowledgments}
This work was supported  by the Romanian National Authority for Scientific
Research through Grant No. IDEI-1012/2011 
for the University of Bucharest.


\begin{thebibliography}{40}
\bibitem{Fol5} These composition rules, albeit general, have the same structure 
as the specialized formulae written in  our Eq.\ (\ref{rhorho}). However, 
they are equivalent to those of the oscillator semigroup treated in Chapter 5 
of Ref.\cite{Fol}.
\end{thebibliography}
\end{document}